\documentclass[epj]{svjour}
\usepackage{graphicx,bm}
\begin{document}
\title{Two-step flux penetration in classic antiferromagnetic superconductor}
\author{T. Krzyszto\'n \and K. Rogacki}
\institute{Institute of Low Temperature and Structure Research,
Polish Academy of Sciences, 50-950 Wroc\l{}aw, P.O.Box 1410,
Poland}
\date{Received:~~~~~~~~\ Revised version:   }
\abstract{The influence of antiferromagnetic order on the mixed
state of a superconductor may result in creation of spin-flop
domains along vortices. This may happen when an external magnetic
field is strong enough to flip over magnetic moments in the vortex
core from their ground state configuration. The formation of
domain structure causes modification of the surface energy
barrier, and creation of the new state in which magnetic flux
density is independent of the applied field. The modified surface
energy barrier has been calculated for parameters of the
antiferromagnetic superconductor DyMo$_{6}$S$_{8}$. The prediction
of two-step flux penetration process has been verified by precise
magnetization measurements performed on the single crystal of
DyMo$_{6}$S$_{8}$ at milikelvin temperatures. A characteristic
plateau on the virgin curve $B(H_0)$ has been found and attributed
to the modified surface energy barrier. The end of the plateau
determines the critical field, which we call the second critical
field for flux penetration. \PACS{
      {74.60.-w}{Type-II superconductivity}   \and
      {74.25.Ha}{Magnetic properties}
     }
} \maketitle

\section*{Introduction}
The discoveries of ternary Rare Earth (RE) Chevrel Phases
REMo$_{6}$S$_{8}$~\cite{Ternary} and RERh$_{4}$B$_{4}$ compounds
with regular distribution of localized magnetic moments of RE
atoms have proved conclusively the coexistence of various types of
magnetism with superconductivity. Intensive experimental and
theoretical works have shown that 4f electrons of RE atoms
responsible for magnetism and 4d electrons of molybdenum
chalcogenide or rhodium boride clusters responsible for
superconductivity are spatially separated and therefore their
interaction is weak. It seems that in many of these systems
superconductivity coexists rather easily with antiferromagnetic
order, where usually the Neel temperature $T_{N}$ is lower than
the critical temperature for superconductivity $T_{c}$. On the
other hand, ferromagnetism and superconductivity cannot coexist in
bulk samples with realistic parameters. Quite often the
ferromagnetic order is transformed into a spiral or domain-like
structure, depending on the type and strength of magnetic
anisotropy in the system~\cite{BulBuzdKulPanj,Maple95}. For almost
two decades the problem of the interaction between magnetism and
superconductivity has been overshadowed by high temperature
superconductivity found in copper oxides. However, the recent
discovery of the presence of magnetic order in Ru-based
superconductors ~\cite{Bauer,Pringle,KlamutX} has triggered a new
series of experiments and inspired a return to the so-called
coexistence phenomenon~\cite{Houzet}. Most recently, the interplay
between magnetism and superconductivity was studied in d-electron
UGe$_{2}$~\cite{Saxena} and ZrZn$_{2}$~\cite{Pfleiderer}, where
itinerant ferromagnetism may coexist with superconductivity, and
in heavy fermion UPd$_{2}$Al$_{3}$~\cite{Sato}, where magnetic
excitons are present in superconducting phase.

\begin{figure}[!htb]
\includegraphics*[width=0.45\textwidth]{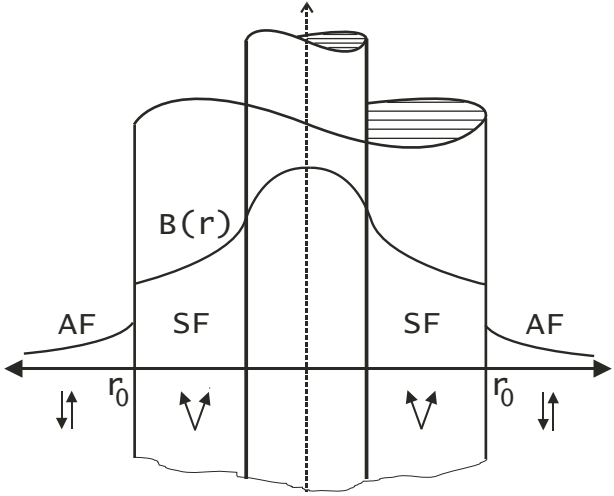}
\caption{The magnetic structure of the isolated vortex and the
distribution of the magnetic induction around the vortex core in
the spin-flop SF and antiferromagnetic AF phases ~\cite{Krzy80}.}
\label{nicstara}
\end{figure}

Among classical magnetic superconductors, the Chevrel phases have
been studied most intensively. These compounds are mainly
polycrystalline materials. However, some specific features can be
measured only on single crystals. One such effect is a two-step
flux penetration process, predicted in Ref.(\cite{Krzy80,Krzy84})
and later observed in an antiferromagnetic superconductor (bct)
ErRh$_{4}$B$_{4}$~\cite{Muto86} and recently in
DyMo$_{6}$S$_{8}$~\cite{Rogacki2001}. This anomaly was explained
as a result of the magnetic transition taking place in the vortex
core. This transition seems to create a new type of vortices with
the unique magnetic structure as shown in Fig.\ref{nicstara}. In
the present paper the two-step flux penetration process is
calculated and measured for a single crystal of DyMo$_{6}$S$_{8}$.
The DyMo$_{6}$S$_{8}$ compound with $T_{c}=1.6~\rm {K}$ exhibits
transition from the paramagnetic to the antiferromagnetic state at
$T_{N}=0.4~\rm {K}$. Its crystal structure can be described as
interconnected Mo$_{6}$S$_{8}$ units and Dy ions. One such unit is
a slightly deformed cube where S atoms sit at the corners and Mo
atoms are situated at the cube-faces. The Mo$_{6}$S$_{8}$ units
are arranged in a simple rhombohedral lattice and Dy ions are
located in the center of the unit cell. The magnetic moments of Dy
ions form a simple structure consisting of $(100)$ planes with
moments of $8.7~\mu _{B}$ alternately parallel and antiparallel to
the $[111]$ rhombohedral axis.

Neutron experiments performed on DyMo$_{6}$S$_{8}$ in an applied
magnetic field at $T=0.2~\rm {K}$ have revealed in the intensity
spectrum a number of peaks characteristic for ferromagnetic
order~\cite{Thomlinson79}. These peaks begin to develop at
$H_{0}=200~\rm {Oe}$, much below the superconducting upper
critical field $H_{c2}$. Thus, in DyMo$_{6}$S$_{8}$ a kind of
ferromagnetic order coexists with superconductivity in the same
manner as antiferromagnetism. For a field applied parallel to the
$[111]$ direction (magnetic easy-axis direction), the
ferromagnetic order is a spin-flop type~\cite{Thomlinson82}. This
feature is easy to understand. Consider the well known phase
diagram of a two-sublattice antiferromagnet. An infinitesimal
magnetic field applied perpendicular to the easy axis makes the
ground antiferromagnetic configuration unstable against the phase
transformation to the canted phase. On the contrary, if the
magnetic field is applied parallel to the easy axis the
antiferromagnetic (AF) phase is stable up to the thermodynamic
critical field $H_{T}$. When the field is further increased, a
spin-flop (SF) phase develops in the system. Let us assume that in
the antiferromagnetic superconductor the lower critical field
fulfills the relation $H_{c1}<\frac{1}{2}H_{T}$ and that the
external field $H_{0}$ is applied parallel to the easy axis. When
$H_{c1}<H_{0}<\frac{1}{2}H_{T}$ the vortices appear entirely in
the AF phase. When $H_{0}$ is increased beyond $\frac{1}{2}H_{T}$
the phase transition to the SF phase originates in the core,
because near $H_{c1}$ the field intensity in the core is
approximately twice $H_{c1}$~\cite{Tinkham}. The spatial
distribution of the field across the vortex is a function
decreasing from the center as seen in Fig.\ref{nicstara}. Thus,
the magnetic field intensity outside the core is less then $H_{T}$
and, therefore, the rest of the vortex remains in the AF phase.
The radius of a spin-flop domain grows as the field is increased.
The formation of domains inside the vortices should be accompanied
by the modification of the surface energy barrier~\cite{Krzy84}.
This process leads to a state of the superconductor in which flux
entrance is temporarily prohibited - flux density is independent
of the applied field. In order to kill this state the external
field should be increased above certain second critical field for
flux penetration. Then, the vortices penetrating the sample will
have the spin flop domains created along the cores. The goal of
our work is to compare the model calculations based on the method
of images ~\cite{ClemLT} with the experimental results obtained
for the DyMo$_{6}$S$_{8}$ single crystal.
\section*{Theoretical considerations}
In order to describe thermodynamic behavior of DyMo$_{6}$S$_{8}$
for constant temperature and constant external magnetic field we
introduce the following free energy
functional~\cite{BulBuzdKulPanj}
\begin{equation}
F=\int dV \{ f_{S}+f_{M}+\frac{1}{8 \pi}( \mathbf{B}-4
\pi\mathbf{M}) ^{2} \} .  \label{eq1}
\end{equation}
Here $\mathbf{B}$ is the magnetic induction, $\mathbf{M}=
\mathbf{M}_{1}+\mathbf{M}_{2}$ the magnetization of two
sublattices antiferromagnet and $\mathbf{H}=\mathbf{B}-4 \pi
\mathbf{M}$ the intensity of thermodynamic magnetic field. The
energy density of the superconducting subsystem $f_S$ is expressed
in a standard way
\begin{equation}
f_{S}=\frac{\hbar ^{2}}{2m}\left| \left( \mathbf{\nabla
}-\frac{2ie} {c \hbar }\mathbf{A}\right)\Psi \right| ^{2}+a \left|
\Psi \right| ^{2}+ \frac{1}{2}b \left| \Psi \right|^4 .
\label{eq2}
\end{equation}
The quantity $e,m,c$ denote charge and mass of the electron and
light velocity, respectively. According to experiments the
antiferromagnetic order is practically unaffected by the
appearance of superconductivity, then it is reasonable to neglect
the effect of superconductivity on the exchange interaction in
$F$. This means that both order parameters $\Psi$ and $\mathbf{M}$
are coupled via the vector potential $\mathbf {A}$:
\begin{eqnarray}
&&\nabla\times \mathbf{A}=\mathbf{B}=\mathbf{H}+4 \pi \mathbf{M}  \nonumber \\
&&\mathbf{j}_{s}= \frac{c}{4\pi}\nabla\times\mathbf{H},\label{eq3}
\end{eqnarray}
where $\mathbf{j}_{s}$ denotes a superconducting current. The
antiferromagnetic energy density, which mimics the experimental
results in DyMo$_{6}$S$_{8}$, is given by the following expression
\begin{equation}
f_{M}=J\mathbf{M}_{1}\cdot\mathbf{M}_{2}+K\sum\limits_{i=1}^{2}\left(
M_{i}^{z}\right)^{2}- \left| \gamma
\right|\sum\limits_{i=1}^{2}\sum\limits_{j=x,y,z}( \mathbf{\nabla
} M_{i}^{j}) ^{2} . \label{eq4}
\end{equation}
Here $J$ is the exchange constant between two antiferromagnetic
sublattices, $K$ denotes single ion anisotropy constant, and
$\sqrt{\left|\gamma\right|}$ is the magnetic stiffness length.
Since in the following we analyze the phenomena with
characteristic length-scales much larger then the interatomic
Dy-Dy distance it is justified to omit the gradient term in
$f_{M}$. The components of the total magnetization vector
$\mathbf{M}=\mathbf{M}_{1}+\mathbf{M}_{2},~| \mathbf{M}_{i}|
=M_{0}~(i=1,2)$ have the following form in both sublattices:
$M_{ix}=M_{0}\sin \theta
_{i},~M_{iy}=0,~M_{iz}=M_{0}\cos\theta_{i}$, where $\theta _{i}$
(canted spin angle) is the angle between the magnetization in the
sublattice and the external magnetic field directed along
$z$-axis. The AF $(\theta _{1}=0,\theta _{2}=\pi )$ and SF phases
$(\theta_{1}=-\theta _{2}=\theta )$ are in thermodynamic
equilibrium in an applied field equal to the thermodynamic
critical field.
\begin{equation}
H_{T}=M_{0}[K(J-K)]^{1/2} . \label{eq5}
\end{equation}
The canted spin angle of the SF phase is then expressed as
\begin{equation}
\cos \theta=\frac {KM_{0}}{H_{T}} . \label{eq6}
\end{equation}
The equilibrium conditions of the whole system can be obtained via
minimization the Gibbs free energy functional
$G=F-\displaystyle\frac{1}{4\pi}\int(\mathbf{B}\cdot\mathbf{H}_{0})dV$
with respect to $\Psi $, $\mathbf{A}$ and $\mathbf{M}$. Performing
this task in London approximation one can obtain
\begin{equation}
\mathbf{B}+\lambda^{2}\nabla\times\nabla\times(\mathbf{B}-4\pi\mathbf{M})=0
, \label{eq7}
\end{equation}
where $\lambda$ is the London penetration depth. The appropriate
equations describing spatial distribution of $\mathbf{M}$ should
accompany Eq.(\ref{eq7}). To make the problem simpler we suppose
that the magnetization is constant across the SF
domain~\cite{Krzy84,Buzdin}. In this way the distribution of the
magnetization around a single vortex is the following
\begin{equation}
|\mathbf{M}| =\left\{
\begin{array}{ccc}
M & ~~if~ & r\leq r_{0} \\
0 & ~~if~ & r>r_{0}
\end{array}
\right. , \label{eq8}
\end{equation}
where $r_{0}$ is the radius of the spin-flop domain. With the help
of Eq.(\ref{eq7}) one can write Eq.(\ref{eq1}) for a single vortex
as follows
\begin{eqnarray}
F&=&\frac{1}{8\pi}\int \left\{ \left(\mathbf{b}_\mathrm{
SF}-4\pi\mathbf{M}\right)^{2}+\lambda^{2}
\left[\nabla\times\left(\mathbf{b}_\mathrm{SF}-4\pi\mathbf{M}\right)\right]^{2}
\right\} dV_\mathrm{SF} \nonumber \\
&+&\frac{1}{8\pi}\int\left[\mathbf{b}_\mathrm{AF}^{2}+\lambda^{2}\left(\nabla\times
\mathbf{b}_\mathrm{AF}\right)^{2}\right]dV_\mathrm{AF} .
\label{eq9}
\end{eqnarray}
Here $\mathbf{b}_\mathrm{ AF}$ and $\mathbf{b}_\mathrm{ SF}$
denote magnetic induction in AF and SF phases of a single vortex,
respectively. The integrals are performed over the volume of each
phase with the exclusion of the volume of the vortex core.
Equation (\ref{eq7}), for a single vortex, can be solved in the
cylindrical coordinates in terms of the modified Bessel functions
$K_{0}$ and $I_{0}$:
\begin{eqnarray}
b_\mathrm{ SF} &=&  C_{1}K_{0}\left(\frac{r}{\lambda}\right)
+C_{2}I_{0}\left(\frac{r}{\lambda}
\right)~,~for~\xi <r\leq r_{0}  \nonumber \\
b_\mathrm{ AF} &=&
C_{3}K_{0}\left(\frac{r}{\lambda}\right)~,~for~r>r_{0} ,
\label{eq10}
\end{eqnarray}
($\xi $ denotes the coherence length) with the following boundary
conditions:
\begin{eqnarray}
b_\mathrm{ SF}\left(\frac{r_{0}}{\lambda}\right) &=& H_{T}+4\pi M = B_{T}\nonumber\\
b_\mathrm{ AF}\left(\frac{r_{0}}{\lambda}\right) &=& H_{T}.
\label{eq11}
\end{eqnarray}
These conditions, together with the flux quantization condition,
are used to calculate the arbitrary constants in Eq.(\ref{eq10}).
\begin{eqnarray}
C_{1} &=&\beta \left[ B_{T} \frac{r_{0}}{\lambda} I_{1}\left(
\frac{r_{0}}{\lambda }\right)
- \alpha I_{0}\left(\frac{r_{0}}{\lambda }\right)\right] \nonumber \\
C_{2} &=& \beta\left\{ B_{T}\left[ \frac{r_0}{\lambda} K_{1}\left(
\frac{r_{0}}{\lambda}
\right) -1\right] + \alpha K_{0}\left( \frac{r_{0}}{\lambda}\right)\right\} \nonumber \\
C_{3}&=&\frac{H_{T}}{K_{0}\left(\displaystyle \frac{r_{0}}{\lambda }\right)}\nonumber \\
\alpha &=& {H_{T}\frac{r_{0}}{\lambda }\frac{K_{1}\left(
\frac{r_{0}}{\lambda }\right)}
{K_{0}\left( \frac{r_{0}}{\lambda }\right)}-\frac{\varphi _{0}}{2\pi \lambda ^{2}}}\nonumber \\
\beta &=& \left\{\displaystyle\frac{r_{0}}{\lambda }K_{1} \left(
\frac{r_{0}}{\lambda }\right) I_{0}\left( \frac{r_{0}}{\lambda
}\right) -I_{0}
\left( \frac{r_{0}}{\lambda }\right)\right.+ \nonumber \\
 &+& \left.\frac{r_{0}}{\lambda }K_{0}\left( \frac{r_{0}}
{\lambda }\right)I_{1}\left( \frac{r_{0}}{\lambda
}\right)\right\}^{-1}. \label{eq12}
\end{eqnarray}
Finally, the minimum of the free energy of the vortex per unit
length
\begin{eqnarray}
\varepsilon _{1}&=&\frac{\lambda ^{2}}{8\pi}\oint_{\sigma
_{1}}d\bm{l} \left\{\left[\textbf{b}_\mathrm{SF}-4\pi\textbf{M}
\right]\times \nabla\times \textbf{b}_\mathrm{SF}\right\}\nonumber\\
&+&\frac{\lambda^{2}}{8\pi}\oint_{\sigma_{2}}d \bm{l}\left\{
\textbf{b}_\mathrm{AF} \times \nabla \times\textbf{b}_\mathrm{AF}
\right\}, \label{eq13}
\end{eqnarray}
with respect to $r_{0}$ determines:
\begin{equation}
\left( \frac{r_{0}}{\lambda }\right) ^{2}=\frac{\varphi
_{0}}{\pi\lambda ^{2}B_{T}}. \label{eq14}
\end{equation}
The line integrals in Eq.(\ref{eq13}) are performed over the
cross-sections perpendicular to the axis of an appropriate
cylindrical element of the surface of the vortex, $\sigma _{1}$
denotes the surface of the vortex core, $\sigma _{2}$ the surface
of the SF domain.

In order to study the conditions under which magnetic flux density
in the sample becomes unstable in the applied magnetic field one
must take into account the surface energy barrier preventing
vortices from entering or exiting the sample. The presence of a
surface of the superconductor leads to the distortion of the field
and current of any vortex located within a distance of the order
of penetration depth from the surface. The requirement that the
currents cannot flow across the surface of the superconductor
leads to the introduction of an image vortex, at $x=-x_{L}$, with
vorticity opposite to the real one. Both vortices interact as real
ones except that the interaction is attractive.

We consider semi-infinite specimen in the half space $x\geq 0$,
the vortex and the external magnetic field running parallel to the
surface. In the low flux density regime $\xi ^{2}<\varphi
_{0}/B<\lambda ^{2}$, Clem ~\cite{ClemLT} has shown the existence
of a vortex-free region of the width $x_{vf}$ near the surface of
the sample and constant vortex density region for $x>x_{vf}$.
Within vortex-free area one can introduce locally averaged
magnetic field $B_{M}$ exponentially decreasing from its surface
value $H_{0}$ to its average interior value $B$,
\begin{equation}
B_{M}=B\cosh \left(\frac{x_{vf}-x}{\lambda }\right). \label{eq15}
\end{equation}
The boundary condition $B_M(0)= H_0$ determines the thickness of
the vortex-free region
\begin{equation}
x_{vf}=\lambda \cosh^{-1}  \left(\frac{H_{0}}{B}\right).
\label{eq16}
\end{equation}
Now we can characterize the distribution of the magnetic induction
around a single vortex in the vortex-free region
\begin{eqnarray}
B_\mathrm{SF}=b_\mathrm{SF}\Big(\frac{x-x_{L}}{\lambda }\Big) &-&
b_\mathrm{ AF}\Big(
\frac{x+x_{L}}{\lambda}\Big)+\nonumber \\
 &+&B_{M}\Big(\frac{x_{vf}-x}{\lambda }\Big), \nonumber \\
B_\mathrm{AF}=b_\mathrm{AF}\Big(\frac{x-x_{L}}{\lambda }\Big)&-& b_\mathrm{AF}
\Big(\frac{x+x_{L}}{\lambda}\Big)+ \nonumber \\
 &+& B_{M}\Big(\frac{x_{vf}-x}{\lambda }\Big).
\label{eq17}
\end{eqnarray}
The Gibbs free energy of the system can now be written in the
following way
\begin{eqnarray}
G&=&\frac{\lambda ^{2}}{8\pi}\oint_{\sigma _{1}}d
\bm{\sigma}\left\{\left[ \mathbf{B}_\mathrm{
SF}-2\mathbf{H}_{0}-4\pi\mathbf{M}\right]
\times\nabla\times \mathbf{B}_\mathrm{SF}\right\}\nonumber\\
&+&\frac{\lambda ^{2}}{8\pi}\oint_{\sigma _{2}}d \bm{\sigma}
\left\{\left[\mathbf{B}_\mathrm{AF}-2\mathbf{H}_{0}\right]
\times\nabla\times \mathbf{B}_\mathrm{AF}\right\}\nonumber \\
&+&\frac{\lambda ^{2}}{8\pi}\oint_{\sigma _{3}}d \bm{\sigma}
\left\{\hat{z}B_{M}\times\nabla\times
\mathbf{B}_\mathrm{AF}\right\}, \label{eq18}
\end{eqnarray}
where $\sigma _{3}$ is the surface of the specimen. After some
transformations~\cite{Krzy84,ClemLT}, one can obtain the Gibbs
free energy per unit length $\mathcal{G}$ in the following form:
\begin{eqnarray}
\mathcal{G} &=&\varepsilon_{1}-\frac{\lambda
^{2}}{4}D_{1}b_{AF}\left(\frac{2x_{L}}{\lambda }\right)\nonumber \\
&-&\frac{\lambda ^{2}}{2}\left[ D_{1}H_{0}-D_{2}B_{M}\left(
\frac{x_{vf}-x}{\lambda }\right) \right], \label{eq19}
\end{eqnarray}
where
\begin{eqnarray}
D_{1}= &-&\left. \xi \frac{\displaystyle
db_\mathrm{SF}\left(\frac{x-x_{L}}{\lambda} \right)}{dx}
\right|_{x=x' } -\left. r_{0}\frac{\displaystyle
db_\mathrm{SF}\left(\frac{x-x_{L}}{\lambda }\right)}{dx}
\right|_{x=x''}\nonumber \\
&-&\left. r_{0}\frac{\displaystyle
db_\mathrm{AF}\left(\frac{x-x_{L}}{\lambda }\right)}{dx}
\right|_{x=x''}\nonumber \\
D_{2}= &-&\left. \xi\frac{\displaystyle
db_\mathrm{SF}\left(\frac{x-x_{L}}{\lambda }\right)}{dx}
\right|_{x=x'} -\left. r_{0}\frac{\displaystyle
db_\mathrm{SF}\left(\frac{x-x_{L}}{\lambda }\right)}{dx}
\right|_{x=x''}\nonumber \\
&-&2\left. r_{0}\frac{\displaystyle
db_\mathrm{AF}\left(\frac{x-x_{L}}{\lambda}\right)}{dx}
\right|_{x=x''}\nonumber
\end{eqnarray}
\[x'=x_{L}+\xi~;~x''=x_{L}+r_{0}\]
$\mathcal{G}$ has its maximum at $x=x_{max}$ somewhere in the
vortex-free region $r_{0}<x_{max}<x_{vf}$. We can find $x_{max}$
solving the force balance equation. When the external field
reaches
\begin{equation}
H_{en2}\left( B\right) =B\cosh \left( \frac
{x_{en}}{\lambda}\right), \label{eq20}
\end{equation}
where $x_{en}$ is the vortex-free width corresponding to an
external field equal to $H_{en2}$, the energy barrier moves within
a distance $r_{0}$ of the surface $(r_{0}\ll x_{vf})$. Thus, one
can get
\begin{equation}
-\frac{\lambda D_{1}}{2D_{2}}\left. \frac{\displaystyle
db_{AF}\left(\frac{2x_{L}}{\lambda} \right)}{dx_{L}}\right|
_{x_{L}=r_{0}}=B\sinh\left( \frac{x_{en}-r_{0}}{\lambda }\right) .
\label{eq21}
\end{equation}
The left hand side of the above equation gives $H_{en2}(0)$. This
field may be thought as the second critical field for flux
penetration calculated in the single vortex
approximation~\cite{Krzy84}.
\begin{equation}
2H_{en2}(0)=\frac{H_{T}}{\displaystyle\sqrt{\frac{\varphi
_{0}}{\pi \lambda ^{2}B_{T}}} \ln \left(\frac{\pi \lambda
^{2}B_{T}}{\varphi _{0}}\right)} . \label{eq22}
\end{equation}
Taking into account that $r_{0}\ll x_{en}$ we finally obtain
\begin{equation}
H_{en2}(B)=\sqrt{B^{2}+H_{en2}^{2}(0)} . \label{eq23}
\end{equation}
Let us make a short summary of the calculations. When the SF
domain develops, the screening current must redistribute its flow
around the vortex in order to fulfill the single flux quantum
requirement. This one can easy deduce from Eqs
(\ref{eq10})-(\ref{eq12}). The redistribution of the screening
current causes the change in the surface energy barrier preventing
vortices from entering into the sample. This is expressed in
Eq.(\ref{eq19}). Consequently, the averaged flux density in the
sample $B=n\varphi_{0}$ remains constant when the external field
is increased. The vortices start to penetrate into the sample
again when the second critical field for flux penetration,
calculated in Eq.(\ref{eq23}), is reached.
\section*{Experimental details}
The single crystals of DyMo$_{6}$S$_{8}$ were grown using the slow
cooling of a melted charge closed in hermetically sealed
molybdenum ampoules. Details of the crystal grow procedure are
described elsewhere~\cite{Rogacki2001,Horyn}. The crystals were
pure, homogeneous and large enough to be used for studying some
subtle effects accompanying the magnetization process at
milikelvin temperatures. Chemical composition and crystal
uniformity were examined using a Hitachi Scanning Electron
Microscope equipped with an energy dispersive x-ray analyzer.
Single-crystal x-ray diffraction measurements were performed at
room temperature on a Simens SMART CCD diffractometer. The
electron probe microanalysis of the regular-shaped crystals showed
a composition corresponding to the DyMo$_{6}$S$_{8}$ formula. The
cell parameters in the rhombohedral lattice were $a_{R}=6.452\cdot
10^{-8}~\mathrm{cm}$ and $\alpha _{R}=89.50^{o}$, and were
equivalent for all crystals analyzed. The single crystal selected
for our experiment had dimensions $0.2\times 0.2\times 0.2\
\mathrm{mm}^{3}$ and a mass $\simeq 0.05~\mathrm{mg}$.

Magnetization was measured with the SHE 330X series SQUID system
with SQUID sensor installed in the vacuum chamber of the $^{3}$He
--$^{4}$He dilution refrigerator. The sensor was thermally
anchored to the liquid He bath $(4.2~\mathrm{K})$ and shielded
with a Nb tube. Two counter-wound pickup coils were connected to
the input coil of the SQUID sensor. The SQUID pickup coils were
placed in the center of a $10$ cm long superconducting solenoid
generating a magnetic field up to $1.5~\mathrm{ kOe}$. Both the
coils and the solenoid were fixed to the mixing chamber of the
dilution refrigerator. Details of the experimental setup are
described elsewhere~\cite{Rogacki2001}. The perfect shielding
$(4\pi\mathcal{M}=H_0)$ of the Meissner state was used to
calibrate the SQUID system. The crystal was oriented with the
magnetic easy axis (the $[111]$ crystallographic triple axis)
parallel to the external magnetic field. For this orientation, the
demagnetizing factor was assumed to be $k=1/3$.
\section*{Comparison with theory}
\begin{figure}[!htb]
\includegraphics*[width=0.45\textwidth]{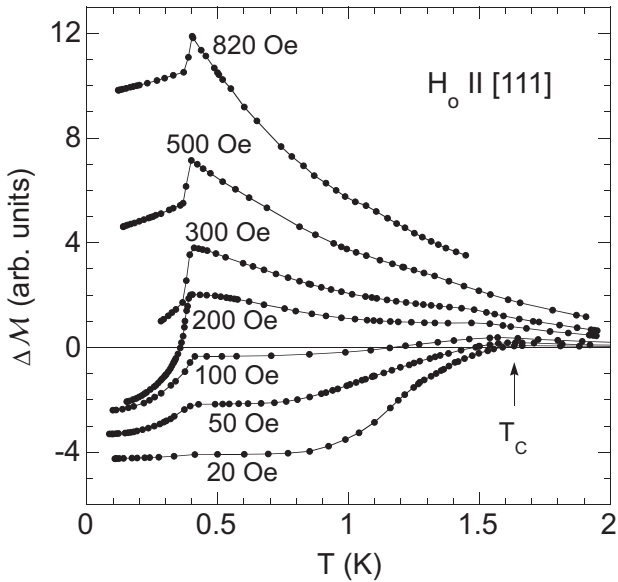}
\caption{Magnetization versus temperature at several applied
fields for DyMo$_6$S$_8$ single crystal for the field direction
oriented parallel to the magnetic easy axis.} \label{magn}
\end{figure}

In Fig.\ref{magn}, the magnetization $\mathcal M$ measured as a
function of temperature is presented for several applied magnetic
fields oriented parallel to the easy axis of the single crystal.
At higher temperatures, the transition to the superconducting
state is observed at $T_{c}$ as the smooth decrease of $\mathcal
M$ (e.g., $T_{c}=1.62~\mathrm{K}~for~H_0=20~\mathrm{Oe}$). This
critical temperature is clearly field dependent as expected for a
superconductor. At low fields, $\mathcal M$ riches negative values
close to $T_{c}$. At higher fields, this is not possible because
of the induced strong paramagnetic moment of the Dy ions. At low
temperatures, the abrupt change of $\mathcal M$ is observed at
$T_{N}=0.4~\mathrm{ K}$,reflecting the transition to the AF state.
In that state, the internal field is reduced and $\mathcal M$ can
now become negative even for higher fields. At $T_{N}$ and for
$H_0\leq 200~\mathrm{Oe}$, the change of $\mathcal M$ between the
paramagnetic (PM) and AF states increases significantly with
increasing field, as expected. However, for $H_0>200~\mathrm{Oe}$,
the single crystal is in the SF phase
~\cite{Thomlinson79,Thomlinson82} and the observed change of
$\mathcal M$, caused by the transition to the ordered state, now
decreases with increasing field.
\begin{figure}[!htb]
\includegraphics*[width=0.45\textwidth]{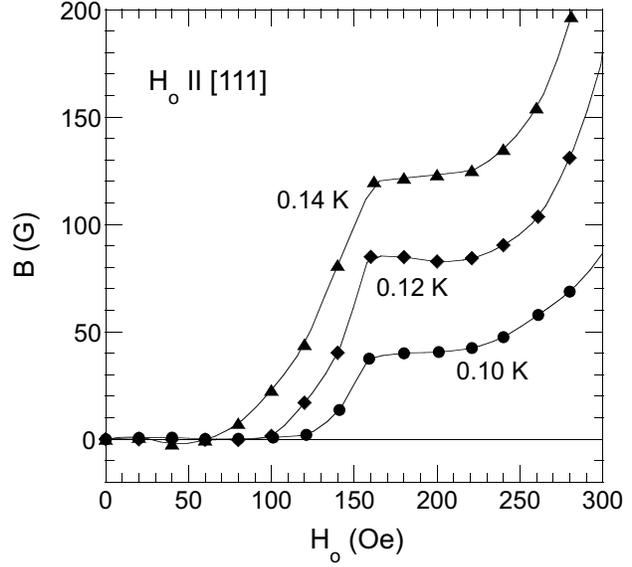}
\caption{Magnetic induction for DyMo$_6$S$_8$ single crystal in
the virgin state measured as a function of an applied field for
three temperatures below $T_N = 0.4~\mathrm{K}$. The field
direction is oriented parallel to the magnetic easy axis of the
crystal. Each $B(H_0)$ curve exhibits characteristic plateau
indicating that a number of vortices is kept constant when the
external field is increased. The results are not corrected for
demagnetizing effects. The corrected values used for calculations
are given in Table~\ref{tab1}. The solid lines are guide to the
eye.} \label{inset}
\end{figure}

The low-field parts of the $B(H_0)$ virgin curves are presented in
Fig.\ref{inset} to show the details of flux penetration. These
curves have been obtained by the simple transformation of the
$\mathcal{M}(H_0)$ results reported in our previous
work~\cite{Rogacki2001}. The observed penetration is typical above
$T_{N}=0.4~\mathrm{K}$ and proceeds as an unusual two-stage
process at lower temperatures where AF order coexists with
superconductivity. At low fields the sample is in the Meissner
state. When the field increases above $H_{c1}$, the sample is
penetrated by the flux. Then, at higher fields, the penetration
process stops unexpectedly and $B=B_{pl}$ is constant in the
sample while the external field is further increased. This new
perfect shielding appears at $H=H_{pl}$. The penetration process
starts again when the field reaches $H=H_{en2}$. This value we
call the second critical field for flux penetration.

An interesting effect of the temperature dependence of $H_{c1}$
below $T_N$ is seen in Fig.3 and Table 1. The very small decrease
in temperature results in the significant increase of $H_{c1}$,
whereas $H_{c2}$ measured by us in $0.1~\mathrm{K}$,
$0.12~\mathrm{K}$, and $0.14~\mathrm{K}$ remains close to
$900~\mathrm{Oe}$. The similar behavior has been observed in
GdMo$_6$Se$_8$, its main feature is a sharp dip on the $H_{c1}(T)$
around $T_N$ and a plateau on $H_{c2}(T)$ for $T<T_N$
~\cite{Rogacki1988}. This behavior agrees well with our
theoretical model. The large increase of $H_{c1}$ for temperatures
decreasing below $T_N$ indicates that our sample is in the
antiferromagnetic collinear phase. In this phase, when temperature
decreases from $T_N$, the pair-breaking effects due to molecular
field and antiferromagnetic fluctuations weaken and $H_{c1}(T)$
rapidly tends to match its pattern in the paramagnetic state. On
the contrary, the nearly constant $H_{c2}$ below
$T_N$~\cite{Ishikawa1982,Ishikawa1978} indicates that the sample
is in the antiferromagnetic canted phase and pair-breaking effects
due to the on-field component of the molecular field are present
even at the lowest temperatures. In our model we have assumed that
vortices enter the sample in the collinear antiferromagnetic
state, and when the field is increased the canted phase appears
first inside the core of the vortex. Thus, above $H_{en2}$, if the
external field is further increased the volume of the canted phase
enlarges. This makes possible to transform the whole sample to the
canted phase well below $H_{c2}$.

In order to compare our theoretical model with the experimental
results we have estimated several quantities. The most important
is the Ginzburg-Landau (GL) parameter $\kappa$. This parameter has
been calculated in~\cite{Rogacki2001} to be equal to $2.6$ for
$T=0.10~\mathrm{K}$. We have taken advantage of the strange
behavior of $H_{c1}(T)$ and $H_{c2}(T)$ to calculate $\kappa$ for
$T=0.14~\mathrm{K}$ and $T=0.12~\mathrm{K}$. The constant value of
$H_{c2}=900~\mathrm{Oe}$ in the range of $0.10~\mathrm{K}~\leq T~
\leq0.14~\mathrm{K}$ predicts that coherence length does not
change in this interval of temperatures where an abrupt increase
of $H_{c1}$ suggests that the penetration depth drastically
decreases. This observation has been used to write the following
equation
\begin{equation}
\frac{H_{c1}(T_0)}{H_{c1}(T)}=\left(\frac{\kappa(T)}{\kappa(T_0)}
\right)^2\frac{\ln\kappa(T_0)}
{\ln\kappa(T)}, \label{eq24}
\end{equation}
where  $T_0=0.10~\mathrm{K}$~,~$\kappa(T_0)=2.6$ and
$0.12~\mathrm{K}~\leq T~ \leq0.14~\mathrm{K}$. The above equation
has been solved numerically and the results for $\kappa$ are given
in the Table~\ref{tab1}.
\begin{table}[h]
\caption{Experimental values of Fig.\ref{inset} corrected for
demagnetizing effects according to the formulae: $H=H_0+4\pi
k\mathcal M$, and $B=H-4\pi (1-k)\mathcal{M}$~\cite{Landau}, where
$\mathcal{M}$ (absolute value) is taken from Fig.4 of
Ref.\cite{Rogacki2001}.} \label{tab1}
\begin{tabular}{l|ccccc}
\hline
$T~[\mathrm{K}]$ &  $\kappa$ & $H_{c1}~[\mathrm{Oe}]$ &  $H_{pl}~[\mathrm{Oe}]$ & $B_{pl}~[\mathrm{G}]$ & $H_{en2}(B)~[\mathrm{Oe}]$\\
 \hline \hline
0.14    &  4.3      & 100           &    170         &    135       &    250 \\
0.12    &  3.1      & 150           &    185         &    105       &    270 \\
0.10    &  2.6      & 180           &    200         &     80       &    280 \\
\hline
\end{tabular}
\end{table}

To find the thermodynamic critical field $H_{T}$ and then to
calculate $H_{en2}(B)$ we have used the following argumentation.
At low fields, in the vicinity of the lower critical field
$H_{c1}$, the intensity of the field in the vortex core is
$2H_{c1}$~\cite{Tinkham}. When the external field is increased the
field intensity in the vortex core increases because of the
superposition of the fields of the surrounding vortices. The field
intensity in the core must reach $H_T$ in order to originate the
transition to the SF phase. Thus, taking into account only the
nearest $z$ neighbors we can write
\begin{equation}
H_{T}=2H_{c1}+z\frac{\varphi _{0}}{2\pi \lambda
^{2}}K_{0}\left(\frac{d}{\lambda}\right) , \label{eq25}
\end{equation}
where $d$ denotes intervortex spacing, and $d/\lambda$ corresponds
to the value $B_{pl}$ for which the penetration process
unexpectedly stops. The relations $B_{\Delta } = 2\varphi
_{o}/d^{2}\sqrt{3}$ (for triangular lattice of vortices), $\varphi
_{o} = 2\pi H_{c2}\xi ^{2}$, where $H_{c2}=900~\mathrm{Oe}$
~\cite{Rogacki2001} have been used to obtain $d/\lambda$. Then,
this value has been inserted into Eq.(\ref{eq25}) to obtain
$H_{T}$. The saturation magnetization of Dy ions, $8\pi M_{0} =
3780~\mathrm{G}$ has been calculated taking into account the
volume of the elementary cell of DyMo$_{6}$S$_{8}$, $V = 268\cdot
10^{-24}~\mathrm{cm}^3$. The anisotropy coefficient
$\mathrm{K}=0.44$ has been determined for each magnetization curve
by finding the best fit of the theoretical with the experimental
magnetization curves~\cite{Morrish}. Next, the magnetization in
the SF-phase domain has been calculated with the help of
Eq.(\ref{eq6})
\begin{equation}
M = 2M_{0}\cos\theta = \frac{2KM_{0}^{2}}{H_{T}}. \label{eq26}
\end{equation}
Eq.(\ref{eq26}) gives $M$ corresponding to the field $H_{pl}$ for
which the penetration stops. Finally, inserting all the above
calculated values into Eqs.(\ref{eq22},\ref{eq23}) we have
obtained $H_{en2}(B)$. The results are summarized in the
Table~\ref{tab2}.
\begin{table}[h]
\caption{Summary of the calculated quantities.} \label{tab2}
\begin{tabular}{l|ccccc}
\hline
$T~[\mathrm{K}]$  & $d/\lambda$ & $H_T~[\mathrm{Oe}]$  & $4\pi M~[\mathrm{G}]$ & $B_T~[\mathrm{G}]$ & $H_{en2}(B)~[\mathrm{Oe}]$\\
\hline \hline
0.14     &   1.7       &  250        &  1000        &  1250     &    215 \\
0.12     &   2.8       &  325        &   775        &  1100     &    240 \\
0.10     &   4.9       &  360        &   700        &  1060     &
265  \\ \hline
\end{tabular}
\end{table}

\section*{Conclusion}
We have demonstrated that the antiferromagnetic superconductor
DyMo$_6$S$_8$ shows interesting behavior in the magnetic field
applied below $T_N$. The sample in the virgin state magnetizes
initially like ordinary type II superconductor. When the applied
magnetic field reaches the critical field for flux penetration the
sample transforms from the Meissner to the mixed state. Then,
magnetization proceeds in an unusual way. As the field is further
increased, a new shielding state appears but, in the contrary to
the Meissner state, with a constant flux density inside the
sample. Characteristic plateau, observed for the magnetization
curves, proofs that magnetic flux density inside DyMo$_6$S$_8$ is
unaffected by the increased external field. When the field reaches
certain value, we call it the second critical field for flux
penetration, the flux starts to enter the sample again. This
phenomenon we name two-step flux penetration. We have argued that
in this new state vortices transform to the shape shown in
Fig.\ref{nicstara}, where a domain of the spin-flop phase is
created. The expected metamorphosis of the vortices leads to a
spatial redistribution of the shielding supercurrents, flowing
around the core, in order to keep constant the flux carried by
each vortex. Consequently, a new energy barrier is formed near the
surface preventing vortices from entering the sample. Thus, the
number of vortices inside the superconductor is kept constant. To
overcome the new energy barrier by the vortices with magnetic
structure the external field must be increased beyond $H_{en2}$,
the second critical field for flux penetration.  The formula for
this field has been derived using the image method. The values of
$H_{en2}$ calculated for three temperatures below $T_N$ agree very
well with the experimental results.
\begin{acknowledgement}
We would like to thank Prof. J. Sznajd, Prof. T. Kope\'{c}, and
Dr. P. Tekiel for helpful discussions. This work was supported by
the State Committee for Scientific Research (KBN) within the
Project \mbox{No. 2 P03B 125 19}.
\end{acknowledgement}


\begin{thebibliography}{99}
\bibitem{Ternary} For review see \textit{Superconductivity in Ternary Compounds},
edited by M. B. Maple, and \O. Fischer, Springer-Verlag, Berlin,
1982.
\bibitem{BulBuzdKulPanj}  L. N. Bulaevskii, A. I. Buzdin, M. Kuli\'{c} and S. V. Panjukov,
Advances in Physics \textbf{34}, 176 (1985), Sov. Phys. Uspekhi
\textbf{27}, 927 (1984).
\bibitem{Maple95}  M. B. Maple, Physica B \textbf{215}, 110 (1995).
\bibitem{Bauer} L. Bauernfeind, W. Widder, and H. F. Braun,  Physica C \textbf{254}, 151 (1995).
\bibitem{Pringle}   D. J. Pringle, J. L. Tallon, B. G. Walker, and H. J. Trodahl,
Phys. Rev. B \textbf{59}, R11679 (1999).
\bibitem{KlamutX}  P. W. Klamut, B. Dabrowski, S. Kolesnik, M. Maxwell, and J. Mais, Phys. Rev. B \textbf{63}, 224512 (2001).
\bibitem{Houzet}  M. Houzet, A. I. Buzdin, and M. Kuli\'{c}, Phys. Rev. B \textbf{64}, 184501 (2001).
\bibitem{Saxena}  S. S. Saxena, P. Agarwal, K. Ahilan, F. M. Grosche, R. K. W. Haselwimmer, M. J. Steiner, E. Pugh, I. R. Walker, S. R. Julian, P. Monthoux, G. G. Lonzarich, A. Huxley, I. Sheikin, D. Braithwaite, and J. Flouquet, Nature \textbf{406}, 587 (2000).
\bibitem{Pfleiderer}    C. Pfleiderer, M. Uhlarz, S. M. Hayden, R. Vollmer, H. v.Lohneysen,
N. R. Bernhoeft, and G. G. Lonzarich, Nature \textbf{412}, 58
(2001).
\bibitem{Sato}   N. K. Sato, N. Aso, K. Miyake, R. Shiina, P. Thalmeier, G. Varelogiannis,
C. Geibel, F. Steglich, P. Fulde, and T. Komatsubara, Nature
\textbf{410}, 340 (2001).
\bibitem{Krzy80}  T. Krzyszto\'n, J. Mag. Mag. Mater. \textbf{15-18}, 1572 (1980).
\bibitem{Krzy84}  T. Krzyszto\'n, Phys. Letters A \textbf{104}, 225 (1984).
\bibitem{Muto86}  H. Iwasaki, M. Ikebe and, Y. Muto, Phys. Rev. B \textbf{33}, 4669
(1986).
\bibitem{Rogacki2001}  K. Rogacki, E. Tjukanoff, and S. Jaakkola,
Phys. Rev. B \textbf{64}, 094520 (2001).
\bibitem{Thomlinson79}  W. Thomlinson, G. Shirane, D. E. Moncton, M. Ishikawa and, \O.
Fischer, J. Appl. Phys. \textbf{50}, 1981 (1979).
\bibitem{Thomlinson82}  W. Thomlinson, G. Shirane, J. W. Lynn, and D. E. Moncton, in
\emph{Superconductivity in Ternary Compounds}, edited by M. B.
Maple, and \O. Fischer, Springer-Verlag, Berlin 1982.
\bibitem{Tinkham}  M. Tinkham, \emph{Introduction to Superconductivity}, chapter 5,
McGraw-Hill Inc., New York 1975.
\bibitem{ClemLT}  J. R. Clem, in \emph{Proceedings of the 13th Conference on Low
Temperature Physics (LT 13)}, vol. 3, Plenum-Press, New York 1974,
p. 102.
\bibitem{Buzdin}  A. I. Buzdin, S. S. Krotov, and D. A. Kuptsov, Solid State Commun.
\textbf{75}, 229 (1990).
\bibitem{Horyn}  R. Hory\'n, O. Pena, C. Geantet, and M. Sergent, Supercond.
Sci. Technol. \textbf{2}, 71 (1989).
\bibitem{Rogacki1988} K. Rogacki, and Cz. Su\l{}kowski, Physica C \textbf{153-155}, 483 (1988).
\bibitem{Ishikawa1982} M. Ishikawa, Contemp. Phys. \textbf{23}, 443-468 (1982).
\bibitem{Ishikawa1978} M. Ishikawa and J. Muller, Solid State Commm. \textbf{27}, 761 (1978).
\bibitem{Landau} L. D. Landau, and E. M. Lifshitz, \emph{Electrodynamics of continuous media}, chapter 6,
Oxford, Pergamon Press 1960.
\bibitem{Morrish} A. H. Morrish, \emph{The Physical Principles of Magnetism}, chapter 6, John Wiley and Sons, Inc. New York 1965.
\end{thebibliography}
\end{document}